\documentclass[journal,reprint,twocolumn]{revtex4-1}
\usepackage{graphicx}
\usepackage{amsmath}
\usepackage{amssymb}
\usepackage{dcolumn}
\usepackage{latexsym}
\usepackage{rotating}
\usepackage[usenames,dvipsnames]{xcolor}
\usepackage{float}
\usepackage{epsfig}
\usepackage{psfrag}
\usepackage{natbib}
\usepackage{bm}
\usepackage{eucal}
\usepackage{braket}
\usepackage{enumerate}
\usepackage{longtable}
\usepackage{subfigure}
\usepackage{bm}
\usepackage{hyperref}
\usepackage{amsfonts}
\usepackage{dcolumn}
\usepackage{bm}
\usepackage{subfigure}

\newcommand{\be}{\begin{equation}}
\newcommand{\ee}{\end{equation}}
\newcommand{\bn}{\begin{eqnarray}}
\newcommand{\en}{\end{eqnarray}}

\usepackage{color} 

\usepackage{hyperref}
\begin{document}

\author{S. Koley$^{1}$}\email{sudiptakoley20@gmail.com}
\thanks{Present Address: Department of Physics, Amity Institute of Applied Sciences, Amity University, Kolkata, 700135, India}
\author{Saurabh Basu$^{2}$}

\title{Superconductivity induced by Ag intercalation in Dirac semimetal  Bi$_{2}$Se$_{3}$}
\affiliation{$^{1}$ Department of Physics, North Eastern Hill University, 
Shillong, Meghalaya, 793022 India}
\affiliation{$^{2}$ Department of Physics, Indian Institute of Technology Guwahati, Assam, 781039 India}
\begin{abstract}
\noindent 
	{\bf In this paper we have investigated the physical and transport properties of 
Bismuth Selenide,(Bi$_2$Se$_3$) intercalated with Silver (Ag) (Gold (Au) is included for comparison)
via dynamical mean field theory with local density approximation. The band 
structure of Bi$_2$Se$_3$ with a Dirac cone is strongly influenced by the 
intercalation phenomena 
at moderate densities which eventually leads to an orbital selective metal 
insulator transition leading to superconductivity at low temperatures. 
The claims on the orbital selectivity are substantiated by computing the 
density of states. While the 
onset of superconducting correlations at low temperatures are supported by a sudden drop 
in the resistivity as a function of temperature and the nature of susceptibility. 
	Further the Fermi surface (FS) maps for low and high temperature yields no FS reconstruction.} 

\end{abstract}


\pacs{
71.10.Hf,
63.20.Dj,
74.72.-h,
74.25.Ha,
76.60.-k,
74.20.Rp
}
\maketitle
\section{INTRODUCTION}
Intercalation is a way of inserting a new material into a lattice without 
significantly changing its parent structure yet modifying its exotic physical 
and transport properties. 
In layered materials intercalation is a well known phenomenon and has good 
applications in a variety of materials, such as, battery electrodes, 
supercapacitors, 
and solid lubricants\cite{wang}. Recent studies have proved that intercalation in topological 
insulators can make the host material superconducting\cite{srxbi}. Search for 
these topological 
superconductors\cite{sato} is a new found interest of condensed matter physics. 
Topological insulators(TI) have insulating bulk states and 
topologically protected metallic surface states. These charge carriers are 
defined
as the Dirac electrons where presence of finite spin-orbital coupling makes the spins of the  
electrons non-degenerate producing a spin-momentum locking effect.

The most renowned example of a 3D TI is Bi$_2$Se$_3$, which has a Dirac cone 
on the surface and a large gap in the bulk \cite{zhang}. The topological 
properties are addressed both theoretically and experimentally by ab initio 
electronic structure 
calculations\cite{zhang} and electron scattering studies\cite{shen}. Bismuth 
selenide is a 
3D layered chalcogenide having a narrow bandgap. Like other layered 
chalcogenides, it also exhibits diverse fascinating electronic properties.
Presence of a van der Waals gap in the crystal structure makes it favorable for acting as a host 
for a number of intercalant materials, such as, Cu, Ag, Sr and many others \cite{koski}. 
Zerovalent intercalation\cite{koski} produces a variety of physical, electronic and magnetic
properties, such as, Rb intercalation\cite{bianchi} that lead to strong Rashba 
spin-orbit coupling (SOC), while intercalating the host material with Sr and Cu 
make it superconducting \cite{srxbi,hor}. Further there are reported experiments 
also for controlling the positions of the Dirac points in Bi$_2$Se$_3$ by 
surface doping\cite{valla}. Also incommensurate and commensurate 
charge density waves are reported in high density intercalation with copper 
in Bi$_2$Se$_3$\cite{koski1}. Thus intercalation introduces new features which 
do not exist in the host material. Moreover the intercalation process has 
renewed interest with 
the prospects of discovering new energy storage materials\cite{ref7-10koski,11,12,13}. The
alkali metals are used for doping surfaces of various topological materials for controlling 
the chemical activity of the surface.
One of the most interesting features of the TI state is the valence and 
conduction bands of different parity cross and subsequently open up a band gap due to 
strong spin orbit interaction, thereby resulting in an inverted band structure 
leading to a Dirac semimetal regime. 
In Bi$_2$Se$_3$ Z$_2$ is the change in parity of the valence band 
eigenvalues at the band-inversion induced time-reversal-invariant point in the 
Brillouin zone.

{\bf With all the experimental and theoretical data cited above, it is needed to 
obtain information on the evolution of structural and electronic 
properties of the alkali atoms intercalated in Bi$_2$Se$_3$.}
It is relevant to mention here that the intercalated compounds are stable; even large
intercalant concentrations do not cause disruptions to the host lattice \cite{koski}.
The parent Bi$_2$Se$_3$, though can be argued as a weak coupling material, however some of the 
earlier experimental and theoretical studies support a strong correlation view\cite{craco}. 
Moreover defect induced Bi$_2$Se$_3$ also shows strong correlation due to the 
interacting pp$\sigma$ bond\cite{mahanti}. 
On the other hand, electron doped Cu$_x$Bi$_2$Se$_3$ shows an effective mass 
enhancement $m^*/m_e$=2.6\cite{sasaki} where $m_e$ is the free-electron mass. So for 
the bulk electronic states in intercalated Bi$_2$Se$_3$ strong coupling route is more appropriate 
in understanding the electronic and transport properties.
Here we use first-principle density functional theory (DFT) 
calculations combined with dynamical mean field theory (DMFT) to show the intercalation
induced changes in the electronic properties of Bi$_2$Se$_3$ with Ag and Au (Au is included for 
comparison) as intercalants. Here we have used $20\%$ intercalation for Ag
throughout this work. Earlier experimental studies \cite{koski} show Ag intercalation,
with concentration greater than $50\%$ exhibits strong satellite peaks
associated with a host Bragg red peak which are
recognized as a signature of an incommensurate charge density wave\cite{wilsonadvphys} 
in superlattice intercalate systems. The positions of the peaks suggest
of a change in the natural periodicity of the Ag intercalated Bi$_2$Se$_3$. 
While Ag intercalation affects the periodicity of the material, experiments 
suggest that Au intercalation only results in crystal defects. Here we shall 
show how they affect the Dirac points and the transport properties of the 
parent Bi$_2$Se$_3$. 

The rest of the paper is structured as follows. A brief description of the 
method, that is, DFT plus DMFT, alongwith the Hamiltonian are outlined in 
section II. The results and discussions appear in section III. Hence we conclude with a summary of our main results
in section IV.

\section{DFT  {\it {plus}} DMFT}
Bi$_2$Se$_3$ has a rhombohedral structure (space group: R3m). The crystal 
structure consists of quintuple layered blocks separated by a van der Waals 
gap, and in each layer, the hexagonal atomic planes are arranged following the 
sequence of Se1-Bi-Se2-Bi-Se1 along the $z$-direction with covalent bonding 
between the atoms. Se1 and Se2 are two inequivalent Selenium atoms. Thus any of the quintuple layers
ends or starts with a Se1 atom and any foreign atom placed in the van der Waals gap will be closer to the
Selenium atom. We add further comments on the structure and its stability in the supplementary information. 

First principles calculations were performed using WIEN2k full-potential 
linearized augmented plane wave (FP-LAPW) ab initio package\cite{pblaha} within 
the DFT formalism\cite{dft} to get the electronic structure and density of 
states (DOS). A $10 \times 10 \times 10$ $k$-mesh (with the cutoff parameter,
R$k_{max}$ = 7.5) is employed here and a generalized gradient approximation 
Perdew-Burke-Ernzerhof (GGA-PBE) exchange correlation potential is
chosen. The muffin-tin radius, R$_{mt}$ is chosen to be 2.5 a.u. for Bi, Se, Au and Ag for 
different intercalation densities. The cell parameters and the (fractional) 
atomic coordinates are derived following earlier experiment\cite{koski}, and 
then these parameters are varied to get
the energy minimized crystal structure in the intercalated Bi$_2$Se$_3$.
Finally the self consistent field (scf) calculations are computed with an 
accuracy of 0.0001 eV.
Thus from the converged scf calculations we have figured out the band structure and the DOS. The band structure and 
atom-resolved DOS is displayed in Fig.1 and Fig.2.  The DOS manifests partially occupied 
Se-4p, Bi-6p and Ag-5s bands (denoted later as a, b, c orbitals respectively) near the Fermi 
level in Ag intercalation, whereas Se-4p, Bi-6p and Au-6s (denoted later as a, b, c orbitals 
respectively) bands near the Fermi level in the Au intercalated Bi$_2$Se$_3$. 
Our results for the parent Bi$_2$Se$_3$ are in accordance with the previous 
calculations\cite{ref16-17zhang2009}. For example, the energy bandgap of about 
0.3 eV which matches nicely with the earlier experimental data. We have 
calculated the parent 
bandstructure both with and without the spin-orbit coupling (SOC). 
Comparing both of them with the existing results, we have concluded that SOC 
induces an
anti-crossing feature around the $\Gamma$ point which indicates inversion of  
the conduction and the valance band, thereby showing Dirac semi-metallic 
nature of Bi$_2$Se$_3$.

For performing electronic and transport calculations, a fully charge-self-consistent 
dynamical mean field theory (DMFT) is used. In a strongly correlated system,
DFT {\it {plus}} DMFT has been successful in explaining a lot of important physics \cite{at1,su}. 
The multi-orbital iterated perturbation theory (MO-IPT), a computationally fast and 
effective impurity solver has been used here. Though not exact, it
works nicely in real systems at both high and low temperature regimes\cite{kotliarrmp}.  
MO-IPT is an approximate method that relies on an interpolation from the
second order perturbation theory for the impurity problem. The interpolation preserves the correct high frequency limit for the
self-energy and is exact in both the non-interacting and
the atomic limits. To present the importance of quantum correlations in an intercalated Bi$_2$Se$_3$ 
we have used multi-orbital Hubbard model with reasonable values for the intra and inter-orbital
Coulomb interactions. The total Hamiltonian is expressed as, 
\begin{equation}
H=\sum_{k,a,b,\sigma}(\epsilon_{k,ab}+\mu_a\delta_{ab})c^{\dagger}_{k,a,\sigma}c_{k,b,\sigma}+U\sum_{i,a}n_{ia\uparrow}n_{ia\downarrow} + $$
$$U'\sum_{i,a,b,\sigma,\sigma'}n_{ia\sigma}n_{ib\sigma'}-J_H\sum_{i,a,b}S_{ia}.S_{ib}
\end{equation}
 where $\epsilon_{k,ab}$ stands for the intra and inter-orbital hopping between three bands which are the hybridization of the p-orbitals
 of Bi, Se and the s orbital of Ag (or Au) that include the effect of SOC also and $U'=U-2J_{H}$ 
where $U$ and $U'$ are the intra- and inter-orbital Coulomb repulsion parameters and $J_H$ is
the Hund’s coupling. Here we have considered $J_H$=0.5 eV and varied $U$ over a range, namely 0.5eV$\le U \le$ 3.5 eV 
(Fig.3). 

\begin{figure*}
\begin{center}
\epsfig{file=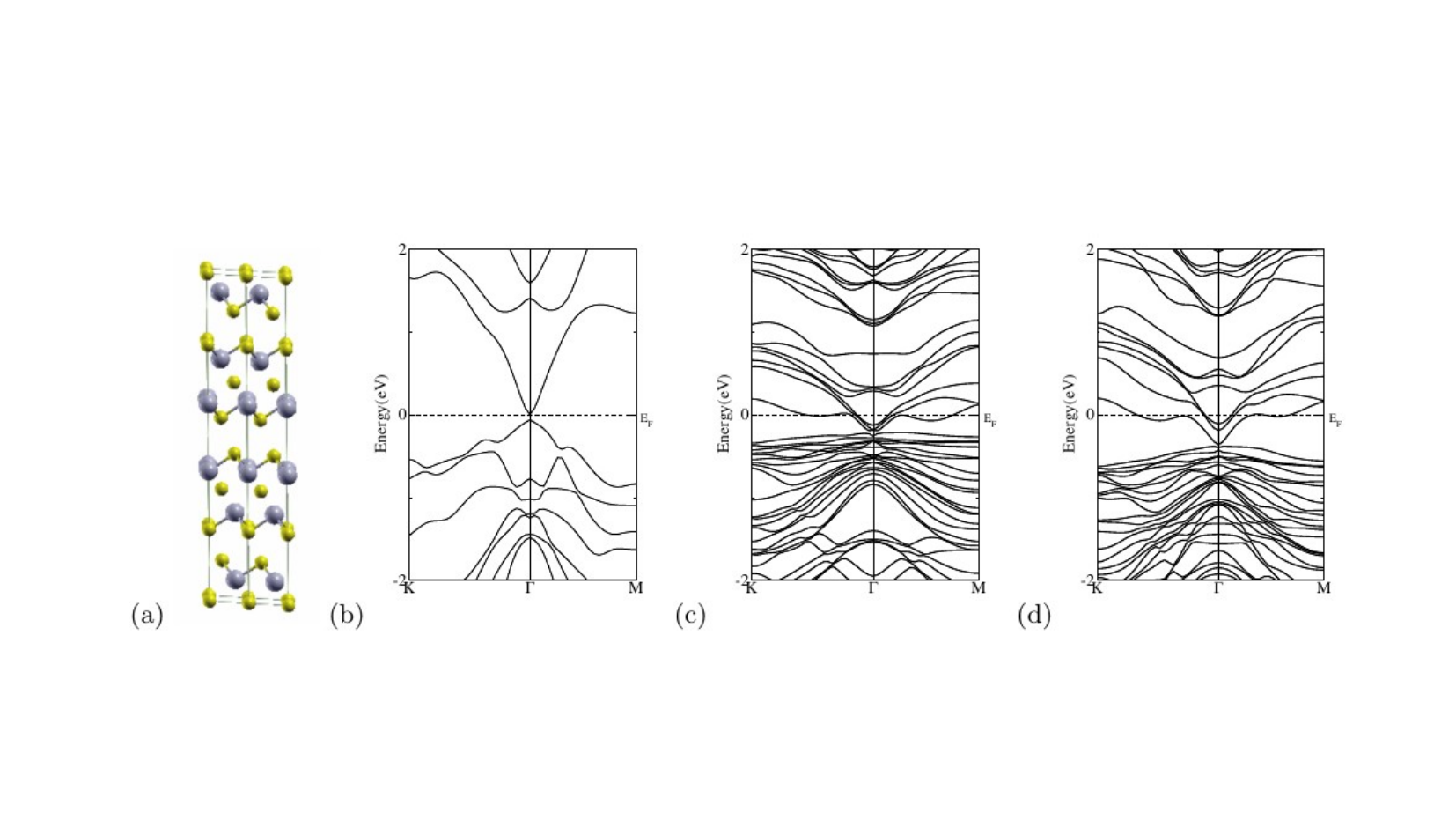,trim=0.8in 1.5in 1.5in 0.3in,clip=false, width=160mm}
\caption{(Color Online) (a) The crystal structure of Bi$_2$Se$_3$. Band structure of
(b) parent Bi$_2$Se$_3$, (c) Ag intercalation and (d) Au intercalation in the
rhombohedral structure including spin-orbit coupling (SOC). Intercalation
results in disappearance of the band inversion at the $\Gamma$ point and results in increase in the
number of conduction electrons. For the effective three-band model
incorporating conduction electrons within DMFT calculations, we use the Bi-p,
Se-p and Ag-5s or Au-6s bands crossing Fermi Energy (E$_F$=0) for DMFT calculation.}
\label{fig1}
\end{center}
\end{figure*}
\begin{figure}
\epsfig{file=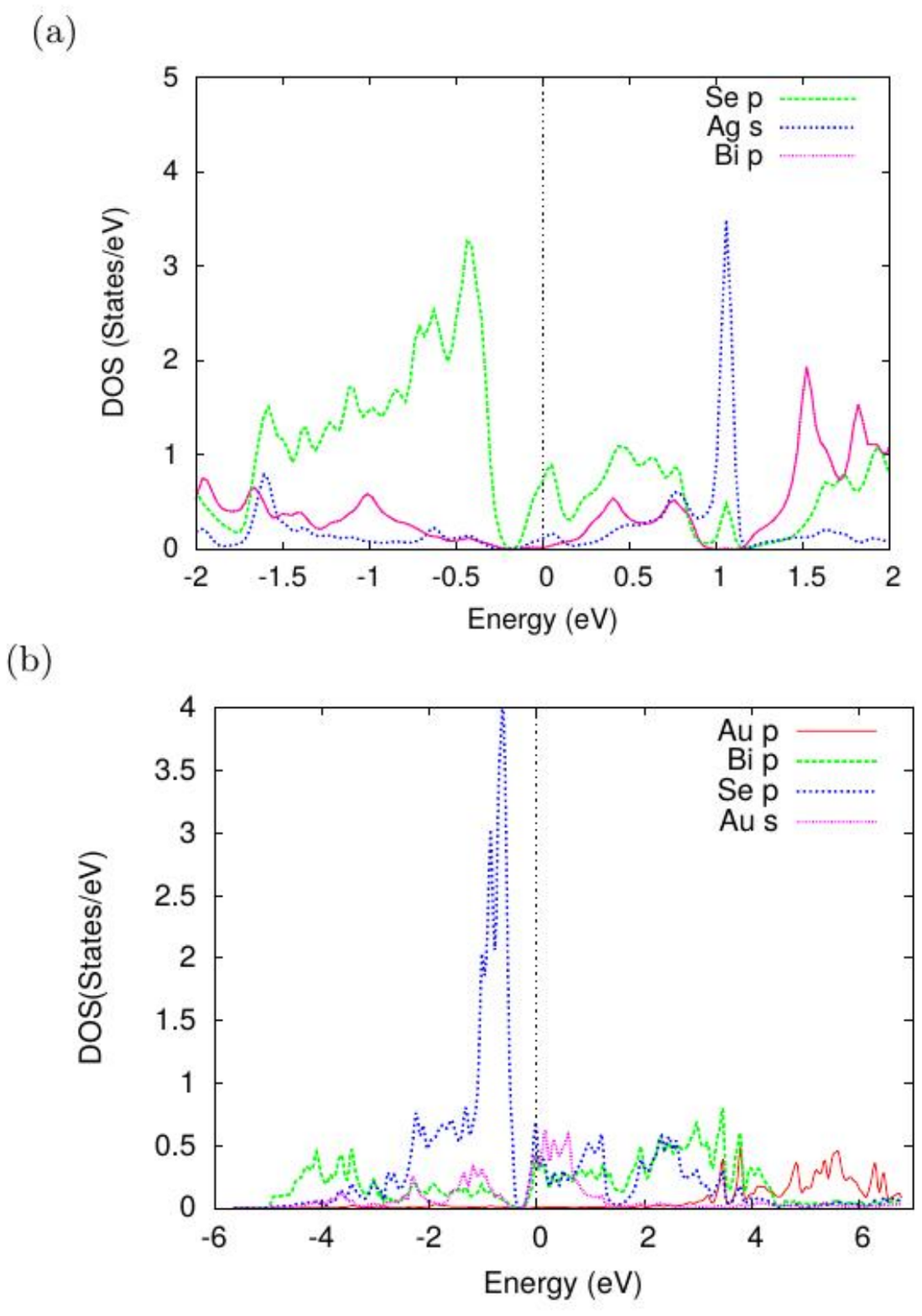,trim=0.0in 0.0in 0.0in 0.0in,clip=false, width=80mm}
\caption{(Color Online) Density of states of the (a) Bi-p, Se-p and Ag-5s and 
(b) Bi-p, Se-p and Au-6s orbitals from DFT calculation.}
\label{fig2}
\end{figure}

\begin{figure*}
\epsfig{file=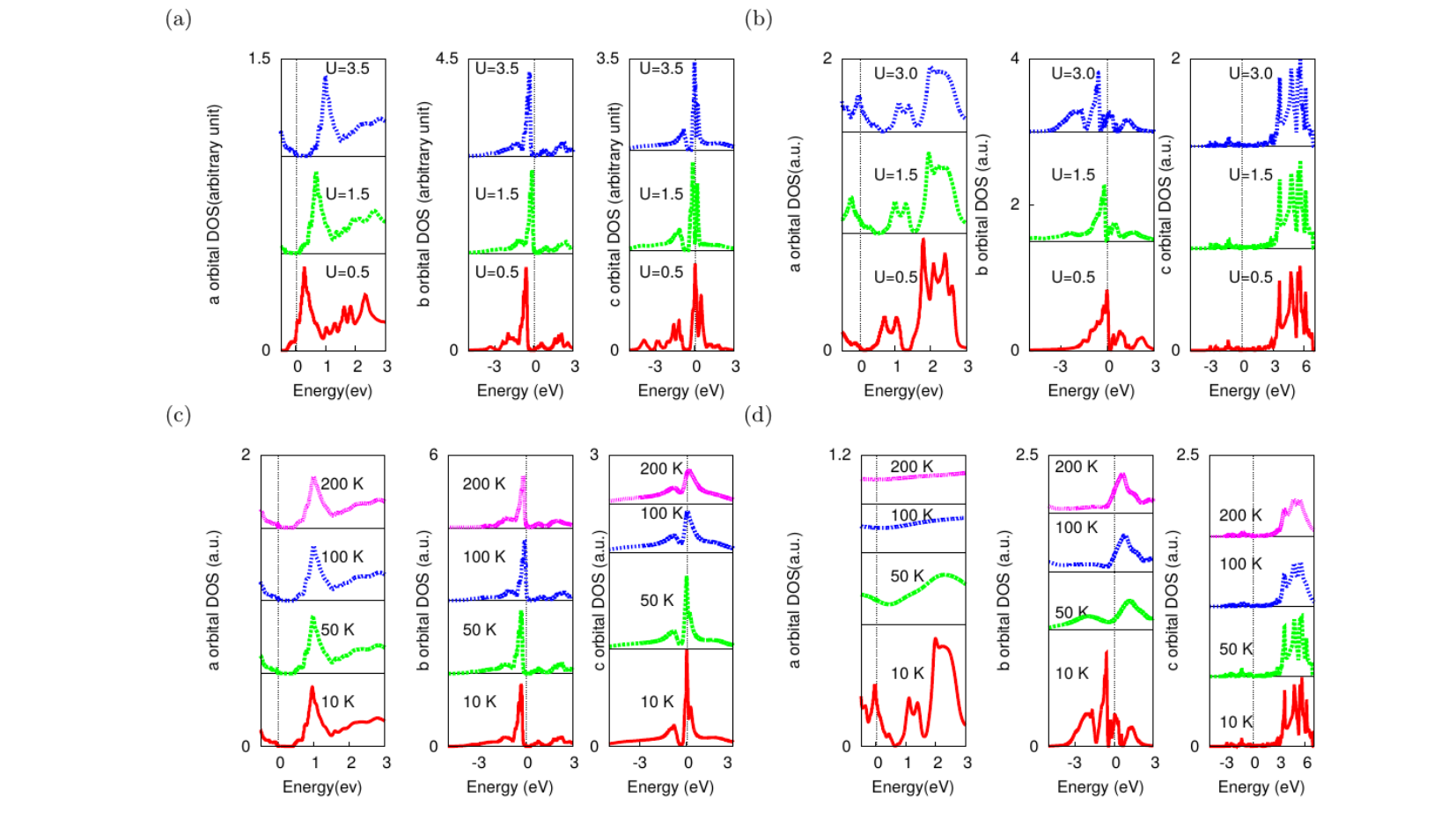,trim=1.0in 0.0in 0.0in 0.0in,clip=false, width=200mm}
\caption{(Color Online) Orbital-dependent density of states from DFT+DMFT at 
various values of $U$ for (a) Ag intercalation and (b)Au intercalation. DFT+DMFT 
density of states at various temperatures for both Ag and Au intercalations
respectively ((c) and (d)). Density of states are shifted in the $y$-axis to 
reflect changes induced by coulomb interaction ((a) and (b)) and due to the inclusion of 
temperature ((c) and (d)). different colors are used for different $U$ and $T$.}
\label{fig3}
\end{figure*}

\section{Results and Discussion}
We shall begin discussing our DFT {\it {plus}} DMFT results. In Fig.3 we present the orbital 
resolved spectral functions of the intercalated Bi$_2$Se$_3$ as a function of the on-site 
Coulomb repulsion, $U$. Intercalation increases the density of conduction 
electrons at the Fermi level. As observed from the DMFT density of states,
with increasing $U$, there is large spectral 
weight transfer in both cases of intercalation with
Ag (Fig. 1c) and Au (Fig. 1d). The spectral weight transfer as a function of 
the Coulomb interaction 
and the low energy quasiparticle resonance are intrinsic to strongly correlated systems. 
The emergence of the Hubbard bands with increasing $U$ is also noteworthy in the intercalated Bi$_2$Se$_3$. For 
the Au intercalated system, the electron-electron interaction only transfers the spectral 
weight and the material remains a correlated metal for all values of $U$,
whereas for the Ag intercalation, increasing $U$ introduces selective gaps in 
the orbital density 
of states which is most prominently seen for $U=3.0$ eV. In this case, the 
large $U$ limit distorts the low energy coherence 
and a selective Mott-Hubbard gap opens up, thereby promoting the formation of 
experimentally
reported charge density wave (CDW) state with a different periodicity. Thus 
existence of  
an orbital selective Mott scenario in the Ag intercalated Bi$_2$Se$_3$ in the 
normal state is related to the strong scattering which might induce CDW order 
with increasing temperature. 

Several interesting features emerge from our results, such as, the 
correlated electronic structure and the orbital dependent spectral weight 
redistribution depending on the strengths $U$ 
and $U'$. The incoherent Hubbard bands are prominent at higher 
energies as observed in the strongly correlated system. While earlier 
theoretical results\cite{craco} for the parent compound, Bi$_2$Se$_3$ have 
reported a Kondo-Mott electronic transition in the
bulk system induced by dynamical correlations and active in all the orbitals,
here we get an orbital selective Mott transition in the Ag intercalated 
compound.

With this orbital selective DOS in Fig.3a (choosing $U=3.0$ eV and $U'=2.0$ eV),
we demonstrate how electron correlations are affected with the inclusion of 
thermal effects. In Fig.3c and Fig.3d we show the temperature 
dependence of Ag and Au intercalated DOS respectively. It is observed that Ag band exhibits a peak 
at the Fermi energy at all values of temperature, while 
there is a pseudogap increasing in the Se-p band with decreasing temperature, 
although the Bi-p band remains almost unaffected. These contrasting behaviors 
are outcomes of multi-orbital electronic correlations on non-interacting DOS. 
While at lower temperatures, the Se-p band remains insulating, but the Bi-p and the Ag-s bands
retain their metallic nature, thereby revealing an interesting physics in these systems. 

Such a scenario can open the possibility of giving rise to novel instabilities 
with competing orders, as those observed in doped 
TiSe$_2$\cite{supc}. Interestingly, at $T=100K$ the orbital selectivity vanishes and a sharp peak appears at the Fermi level from below.
Whereas earlier x-ray diffraction data\cite{koski} show that there is formation of a CDW order in the intercalated compound, 
here we find opening of a pseudogap below 100K. This is a novel finding in these types of systems.

\begin{figure*}
\epsfig{file=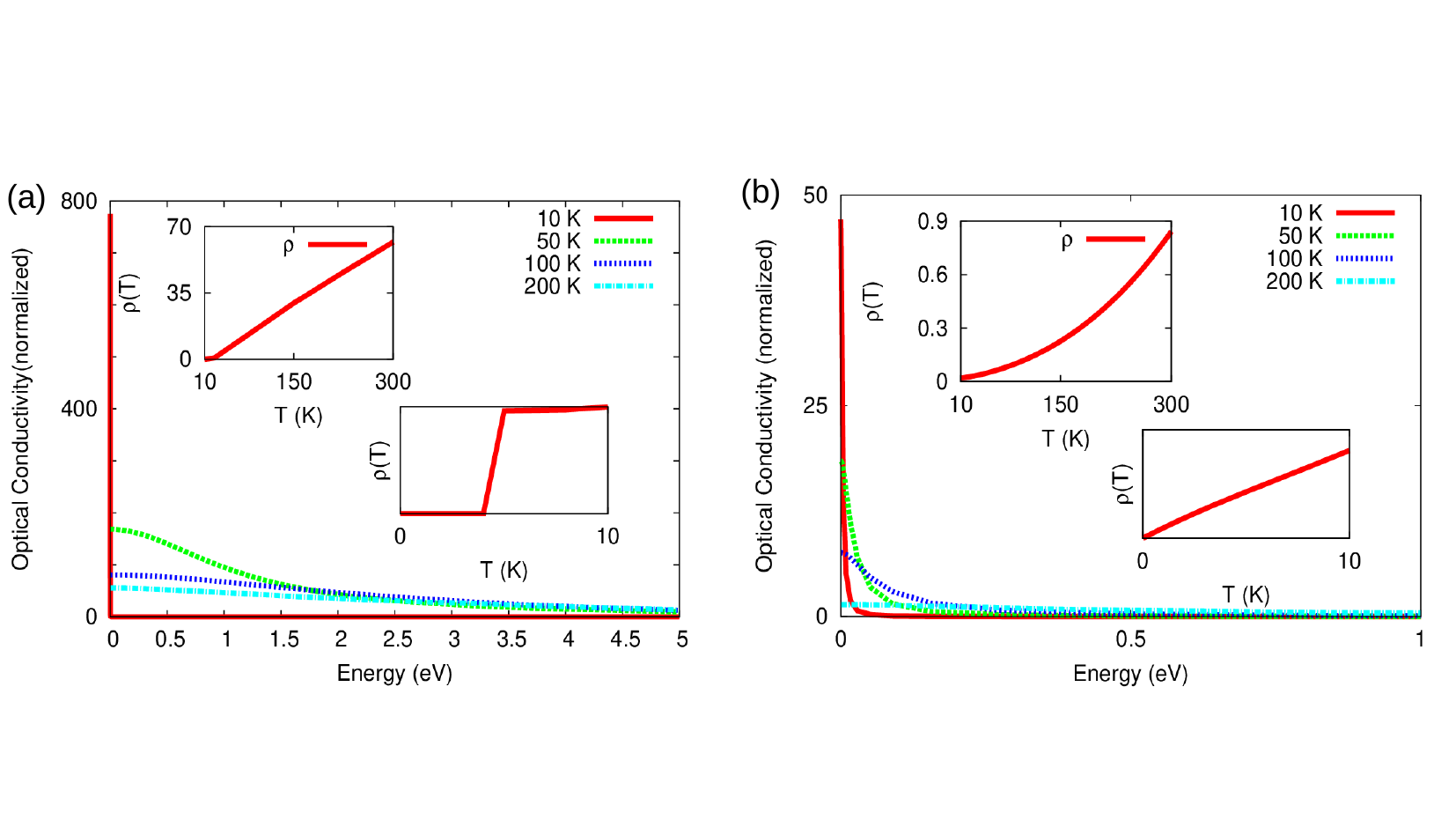,trim=1.0in 0.2in 0.0in 1.0in,clip=false, width=100mm}
\caption{(Color Online) Optical conductivity (normalized with 
respect to normal state value) of (a) Ag and (b) Au intercalated 
Bi$_2$Se$_3$. For Ag intercalation, high temperature optical conductivities (50 K-200K) are 
multiplied by 1000 to show their variation in comparison with 10 K since the corresponding 
values are very small. Inset shows resistivity plot at two different temperature range, namely, 
10K-100K and 0K-10K. For Ag intercalation, the resistivity at low temperature range shows sudden 
decrease below 5K.}
\label{fig4}
\end{figure*}
Further we have computed the optical conductivity in Fig.4 as a function of 
energy for a few different values of temperature, $T$ from the DMFT Green's 
function.
The DMFT equation for optical conductivity is as follows,
\begin{eqnarray*}
\sigma(\omega)=\sigma_{0}\int d\epsilon \rho_{0}(\epsilon) 
\int d\nu \frac{f(\omega+\nu)-f(\nu)}{\omega}\rho_{\epsilon}(\omega+\nu)\rho_{\epsilon}(\nu)
\end{eqnarray*}
where the density of states, $\rho_{\epsilon}(\omega)$ is defined as, 
$\rho_{\epsilon}(\omega)=\sum_{a=1}^{2} \rho_{\epsilon_{a}}(\omega)=(-1/\pi)\sum_{a=1}^{2} 
Im[1/(\omega+\mu_{a}-\epsilon_{a}-\Sigma_{a}(\omega))]$, and $f(\nu)$ is the Fermi function.
In DMFT, the above calculation for optical conductivity is simplified and the transport coefficients can be
computed directly from the DMFT propagators\cite{biermann}. The simplifications originate 
from the fact that the irreducible vertex corrections
vanish for a one-band case and turn out to be surprisingly small corresponding 
to a multiband
case.
In the optical conductivity corresponding to both the systems we have plotted 
the relevant energy range from the band structure for those three bands which 
are crossing Fermi level.
In Ag intercalation, specifically up to an energy of about 0.5 eV, the   
optical conductivity lineshapes show sizable spectral 
weight transfer with increasing temperature. The sharp peak in the 
conductivity at low 
temperature is not a Fermi Liquid (FL) Drude peak, rather this peak arises due 
to reduced incoherence induced by the 
CDW transition.
Clear spectral weight transfer upto 1 eV is discernible from the conductivity
lineshapes, which is a very high value in comparison with the energy range for 
the bands indicating large dynamical correlations. The overall change in the 
optical conductivity points towards appearance of an order. 
In contrast, Au intercalation shows a FL (Im$\Sigma(\omega)\simeq$ 0 as $\omega\simeq$ 0 where $\Sigma(\omega)$ is DMFT self energy) behavior (consistent with 
the DOS data from our DMFT calculations) throughout the temperature range. 
It shows a featureless broad peak 
till about 0.53 eV and an additional Drude peak at lower energies. The low energy peak height decreases 
with temperature, and alongwith a spectral weight transfer occurs at higher energies. 

\begin{figure}
\begin{center}	
\epsfig{file=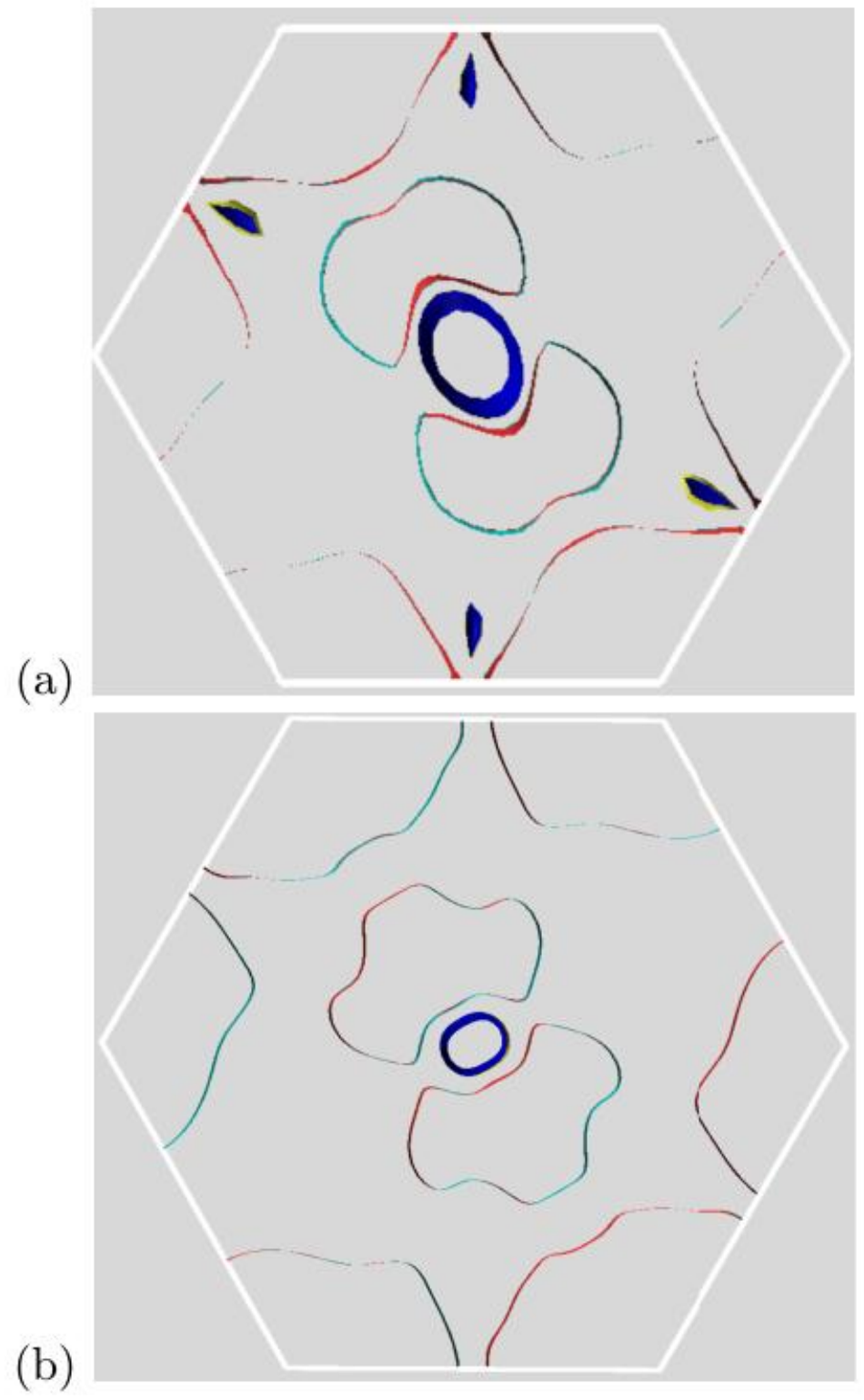,trim=0.9in 0.9in 0.9in 0.7in,clip=false, width=80mm}
\caption{(Color Online) DMFT Fermi surface map for Ag intercalated Bi$_2$Se$_3$ 
at (i) low (10K) and (ii) high(200 K) temperature. The central pocket 
changes its lineshape along with disappearance of pockets at the M points.}
\label{fig5}
\end{center}
\end{figure}

\begin{figure}
\begin{center}	
\epsfig{file=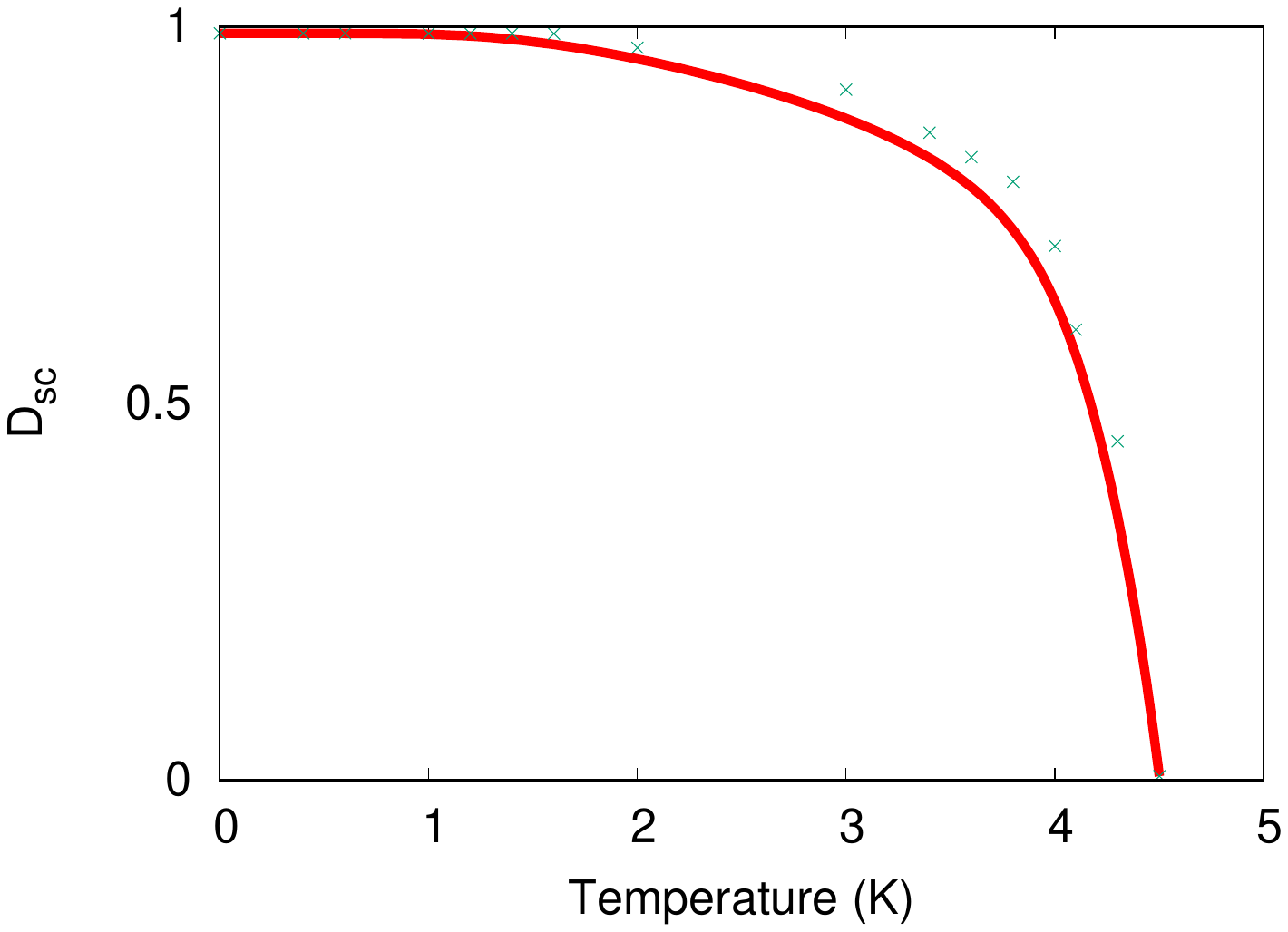,trim=0.9in 0.9in 0.9in 0.7in,clip=false, width=80mm}
\caption{(Color Online) DMFT superconducting order parameter plot with 
temperature. Red line is a fit to the actual data points which are represented 
	by green points.}
\label{figr65}
\end{center}
\end{figure}
We next compute the dc resistivity 
from our dynamical mean field theory. 
The analysis of our resistivity data (inset of Fig.4a) shows linear in $T$ 
behavior of the Ag intercalated compound over a large range of temperature, 
and all the way till room temperature. This
is one of the key results of our paper which allows us to conclude 
that the transition to a ordered phase is a coherence restoring transition 
which occurs at higher temperatures.  
The linear $T$ dependence of resistivity is in good accord with the finding of 
a CDW order in experiments, and hence mandates deeper microscopic exploration. 
However for the Au intercalation, there is no such orbital selectivity and the 
spectral 
function drastically changes at temperatures greater than 10K. All the three 
bands continue to show metallic behavior throughout the temperature range. 
Clearly the case concerning the intercalation of Au differs significantly from 
that of Ag.

Let us look at the low temperature scenario. 
To show the behavior of $\rho(T)$ at lower temperatures, we have plotted it separately 
in the lower inset of Fig.4a in the range 0-10K which shows sharp decrease 
in resistivity (nearly vanishing) occurring between 2K and 5K. This 
provides a theoretical evidence of finding superconductivity in the Ag 
intercalated Bi$_2$Se$_3$. 
The inset of Fig.4b shows the temperature dependent resistivity of Au 
intercalated Bi$_2$Se$_3$. Consistent with findings of the FL behavior, the dc 
resistivity shows $T^2$ dependence throughout the temperature range. Thus, as 
predicted from experiments, Au intercalation only introduces defects in the 
Bi$_2$Se$_3$ crystal lattice which makes the system metallic.

Although Ag is plasmonic, however onset of superconductivity with doping by Ag  
is not new phenomena. It was reported earlier that an enhancement of 
$T_c$ can be achieved with Ag present in a semiconductor matrix\cite{djurek}.

Now to explore the microscopic origin of ordering in
Ag intercalated Bi$_2$Se$_3$, we compute DMFT Fermi surface (FS) (Fig.5) at 
high and low temperature.
In parent Bi$_2$Se$_3$ due to band inversion, there is no Fermi level crossing, thus 
resulting in no pockets in the FS map. Strong spin-orbit interaction eliminates 
the possibility of FS nesting. From the DMFT Fermi surface estimations, the 
presence of selective gaps is observed, though there is no Fermi surface 
reconstruction with temperature, 
which is in accordance with the corresponding features realized for iron 
superconductors and chalcogenides and the superconductivity is originated due to electron electron interaction 
\cite{supc}. 
The two particle interaction, obtained to second order is proportional to 
$\epsilon_{ab}^2$, which is more relevant in ordered low $T$ region. The interaction 
is $H_{res}\simeq -\epsilon_{ab}^{2}\chi_{ab}(0,0)\sum_{<i,j>,\sigma\sigma'}
c_{ia\sigma}^{\dag}c_{jb\sigma}c_{jb\sigma'}^{\dag}c_{ia\sigma'}$,
with $\chi_{ab}(0,0)$ the inter-orbital susceptibility
calculated from the normal state DMFT results. 
Now the new effective Hamiltonian is $H=H_{n}+H_{res}^{HF}$, where
$H_{n} = \sum_{k,\nu}(\epsilon_{k,\nu}+\Sigma_{\nu}(\omega)-E_{\nu})
c_{k,\nu}^{\dagger}
c_{k,\nu}+\sum_{a\ne b,(k)}\epsilon_{ab}( c_{k,a}^{\dagger}c_{k,b}+h.c.)$,
with $\nu=a,b$.
The residual Hamiltonian $H_{res}^{HF}$ is found by 
decoupling the intersite interaction in a generalised Hartree Fock(HF) sense. 
Now this will 
produce ordering in particle-particle (SC) channel. The HF Hamiltonian is  
$H_{res}^{HF}=
-p\sum_{\langle i,j\rangle,a,b,\sigma,\sigma'}(\langle n_{i,a}\rangle
n_{j,b}+\langle n_{j,b}\rangle n_{i,a}-\langle
c_{i,a,\sigma}^{\dagger}c_{j,b,\sigma'}^{\dagger}\rangle c_{j,b,\sigma'}c_{i,a,\sigma}+ h.c.)$ (where p is proportionality constant). We studied 
here the superconducting phase with the two particle instability in 
particle-particle channel (with parametrized p=0.1) and the 
superconducting order parameter 
can be calculated from $D_{sc} \propto \langle c_{i,a,\sigma}^{\dagger}c_{j,b,\sigma'}^{\dagger}\rangle$ which yields multiband spin-singlet SC. 
The anomalous Green's function:
$F(k,\tau) \equiv -\langle T_{\tau}c_{k\uparrow}(\tau)c_{-k\downarrow}$
$(0)\rangle$ 
satisfying $F(-k,-\tau)=F(k,\tau)$ for s wave pairing is introduced in the 
low temperature phase of intercalated Bi$_2$Se$_3$. We presented superconducting order parameter calculated self consistently from DMFT in Fig.6. The order 
parameter follows $(1-T/T_c)^{0.5}$ behaviour and vanishes at 4.5 K where the 
resistivity also drops.\\

The electron-electron interaction induced ordering is supported by 
negligible temperature effects on the Fermi 
surface lineshapes. Now in the Ag intercalation case, high temperature orbital 
selectivity is observed which affects the FS too. The high temperature FS is 
substantially 
smeared out at the M points. Further examination of the DMFT FS map at 
the $\Gamma$ point reveals that
the central pocket acquires an oblong shape, instead of being hexagonal, as is usually  
expected from the DFT results. This change in FS can possibly arise due to the orbital 
dependent reconstruction as a result of the inter-orbital interactions.  
Further at low temperatures, the electron pockets near the $\Gamma$ point change 
their shape. New FS sheets appear at low temperatures at the M point. Since these 
features become well defined at low temperature, that is at temperatures below which 
superconductivity 
is predicted, it should be a direct consequence of superconducting order induced 
reconstruction of electronic states.
 
\section{Conclusions}

To summarize, here we have presented a DFT {\it {plus}} DMFT study of Ag and Au 
intercalated Bi$_2$Se$_3$ to 
explore the ordering induced by these intercalants. The work is focused mainly 
on the changes in electronic and transport properties that occur in this scenario from our
DFT calculations. 
The electronic band structure of parent Bi$_2$Se$_3$ displays
signature of a Dirac semimetal, which substantially changes due to intercalation.
Intercalation with Ag induces an increase in the conduction electron density and 
the band inversion at the $\Gamma$ point disappears. We have also observed an orbital 
selective
Mott-Hubbard gap and studied effects of the interaction parameter therein. Interestingly at low
temperature, superconducting correlations emerge that are convincingly shown by a sudden drop in the 
resistivity below $T_{c} \sim 4.5$K which is further supported by the plot of 
superconducting order parameter which drops to zero at $T_{c}$. Finally the absence of a structural
transition induced is confirmed by the FS maps at both low
and high temperature. The latter suggests that the FS is gapped only in certain directions and hence
invalidates the concept of nesting as the origin of order. Armed with all the DMFT results,
including the resistivity data with an effective three-band model, the onset of superconductivity
is predicted with Ag intercalation. 
This study can be used to other layered materials with similar structure like 
other layer topological compounds emerging as a novel research topic 
for its application in energy storage.

{\bf {Acknowledgement}}
S. Koley acknowledges DST women scientist grant SR/WOS-A/PM-80/2016(G) for finance 
and also thank Prof. M C Mahato for mentoring and useful conversations. SB acknowledges 
financial support from the SERB grant No. EMR/2015/001039.

\end{document}